\documentclass[aps,prd,eqsecnum,preprint,tightenlines,nofootinbib,showpacs]{revtex4-1}
\usepackage{graphicx,amsmath,latexsym}

\def\bea{\begin{eqnarray}}
\def\eea{\end{eqnarray}}
\def\be{\begin{equation}}
\def\ee{\end{equation}}
\newcommand{\ub}[1]{\underline{#1}}
\newcommand{\ob}[1]{\overline{#1}}
\newcommand{\Pminus}{{\cal P}^-}
\newcommand{\veck}{\vec{k}_\perp}
\newcommand{\veckp}{\vec{k}_\perp^{\,\prime}}

\begin{document}

\title{Application of a light-front coupled-cluster method to quantum electrodynamics\footnote{Presented 
at the Sixth International Conference on Quarks and Nuclear Physics,
		 April 16-20, 2012,
		 Ecole Polytechnique, Palaiseau, Paris.  To appear in the proceedings.}}

\author{Sophia S. Chabysheva}
\affiliation{Department of Physics \\
University of Minnesota-Duluth \\
Duluth, Minnesota 55812 USA}

\date{\today}

\begin{abstract}
A field-theoretic formulation of the exponential-operator
technique is applied to a Hamiltonian eigenvalue
problem in electrodynamics, quantized in light-front 
coordinates.  Specifically, we consider the
dressed-electron state, without positron contributions but with an
unlimited number of photons, and compute its anomalous magnetic
moment.  A simple perturbative solution immediately yields the
Schwinger result of $\alpha/2\pi$.  The nonperturbative solution,
which requires numerical techniques, sums a subset of corrections to
all orders in $\alpha$ and incorporates additional physics.
\end{abstract}

\maketitle

\section{Introduction}

Although the nonperturbative light-front coupled-cluster (LFCC)
method~\cite{LFCClett} is intended for strongly coupled theories,
where perturbation theory is of limited use, we explore its
utility in the context of a gauge theory by considering the
dressed-electron state in quantum electrodynamics (QED)~\cite{LFCCqed}.
The method requires the light-front coordinates of Dirac~\cite{Dirac,DLCQreview},
where the Hamiltonian evolves a state along the time
direction $x^+=t+z$.  The spatial coordinates are
$\ub{x}=(x^-\equiv t-z,\vec{x}_\perp\equiv(x,y))$.
The light-front energy conjugate to the chosen time is
$p^-\equiv E-p_z$, and the corresponding light-front
momentum is $\ub{p}=(p^+\equiv E+p_z,\vec{p}_\perp\equiv (p_x,p_y))$.
In these coordinates, the fundamental Hamiltonian eigenvalue
problem is $\Pminus|\psi\rangle=\frac{M^2+P_\perp^2}{P^+}|\psi\rangle$.
Ordinarily, this eigenvalue problem is solved approximately by
a truncated Fock-space expansion of the eigenstate.  The LFCC
method solves the problem without Fock-space truncation
by building the eigenstate as
$|\psi\rangle=\sqrt{Z}e^T|\phi\rangle$ from a valence state $|\phi\rangle$
and an operator $T$ that increases particle number while conserving
any quantum numbers of the valence state.  The constant $Z$ is a
normalization factor.

The valence state is then an eigenstate of an effective Hamiltonian
$\ob{\Pminus}=e^{-T}\Pminus e^T$, which can be computed using
a Baker--Hausdorff expansion
$\ob{\Pminus}=\Pminus+[\Pminus,T]+\frac12 [[\Pminus,T],T]+\cdots$.
The new eigenvalue problem is
$\ob{\Pminus}|\phi\rangle=\frac{M^2+P_\perp^2}{P^+}|\phi\rangle$.
When projected onto the valence and orthogonal sectors, it
becomes
\be
P_v\ob{\Pminus}|\phi\rangle=\frac{M^2+P_\perp^2}{P^+}|\phi\rangle, \;\;\;\;
(1-P_v)\ob{\Pminus}|\phi\rangle=0,
\ee
where $P_v$ is the projection onto the valence sector.

A matrix element such as $\langle\psi_2|\hat O|\psi_1\rangle$ can
be calculated, with $|\psi_i\rangle=\sqrt{Z_i}e^{T_i}|\phi_i\rangle$
and $Z_i=1/\langle\phi_i|e^{T_i^\dagger}e^{T_i}|\phi_i\rangle$.
We define $\ob{O_i}=e^{-T_i}\hat O e^{T_i}
               =\hat{O}_i+[\hat{O}_i,T]+\frac12[[\hat{O}_i,T],T]+\cdots$
and, to avoid the infinite sum in the denominator,
\be
\langle\tilde\psi_i|\equiv\langle\phi|\frac{e^{T_i^\dagger}e^T_i}
      {\langle\phi|e^{T_i^\dagger} e^{T_i}|\phi\rangle}
      =Z_i\langle\phi|e^{T_i^\dagger}e^{T_i}=\sqrt{Z_i}\langle\psi_i|e^{T_i}.
\ee
We then have
\be
\langle\psi_2|\hat O|\psi_1\rangle
    =\sqrt{Z_1/Z_2}\langle\tilde\psi_2|\ob{O_2}e^{-T_2}e^{T_1}|\phi_1\rangle
=\sqrt{Z_2/Z_1}\langle\tilde\psi_1|\ob{O_1^\dagger}e^{-T_1}e^{T_2}|\phi_2\rangle^*,
\ee
and, therefore,
\be
\langle\psi_2|\hat O|\psi_1\rangle
  =\sqrt{\langle\tilde\psi_2|\ob{O_2}e^{-T_2}e^{T_1}|\phi_1\rangle
     \langle\tilde\psi_1|\ob{O_1^\dagger}e^{-T_1}e^{T_2}|\phi_2\rangle^*}.
\ee
In the diagonal case, this reduces to
\be
\langle\psi|\hat O|\psi\rangle
  =\langle\tilde\psi|\ob{O}|\phi\rangle.
\ee
The $\langle\tilde\psi_i|$ can be shown to be left eigenstates of
the effective Hamiltonian.

We apply this to QED in an arbitrary covariant gauge,
for which the Pauli--Villars-regulated Lagrangian is~\cite{ArbGauge}
\bea
{\cal L} &=&  \sum_{i=0}^2 (-1)^i \left[-\frac14 F_i^{\mu \nu} F_{i,\mu \nu} 
         +\frac12 \mu_i^2 A_i^\mu A_{i\mu} 
         -\frac12 \zeta \left(\partial^\mu A_{i\mu}\right)^2\right] \\
&&+ \sum_{i=0}^2 (-1)^i \bar{\psi_i} (i \gamma^\mu \partial_\mu - m_i) \psi_i 
  - e \bar{\psi}\gamma^\mu \psi A_\mu .  \nonumber
\eea
Here the fundamental physical ($i=0$) and Pauli--Villars ($i=1$) fields
appear in null combinations
\be \label{eq:NullFields}
  \psi =  \sum_{i=0}^2 \sqrt{\beta_i}\psi_i, \;\;
  A_\mu  = \sum_{i=0}^2 \sqrt{\xi_i}A_{i\mu}, \;\;
  F_{i\mu \nu} = \partial_\mu A_{i\nu}-\partial_\nu A_{i\mu} .
\ee
The coupling coefficients $\xi_i$ and $\beta_i$ are constrained by
\be
\xi_0=1, \;\;
\sum_{i=0}^2(-1)^i\xi_i=0, \;\;
\beta_0=1, \;\;
\sum_{i=0}^2(-1)^i\beta_i=0.
\ee
To fix $\xi_2$ and $\beta_2$, we
require chiral symmetry restoration in the zero-mass limit~\cite{ChiralLimit}
and a zero photon mass~\cite{VacPol}.
The light-front Hamiltonian, without antifermion terms, is then found to be~\cite{ArbGauge}
\bea \label{eq:QEDP-}
\lefteqn{\Pminus=
   \sum_{is}\int d\ub{p}
      \frac{m_i^2+p_\perp^2}{p^+}(-1)^i
          b_{is}^\dagger(\ub{p}) b_{is}(\ub{p}) 
          +\sum_{l\lambda}\int d\ub{k}
          \frac{\mu_{l\lambda}^2+k_\perp^2}{k^+}(-1)^l\epsilon^\lambda
             a_{l\lambda}^\dagger(\ub{k}) a_{l\lambda}(\ub{k})}&&  \\
   && +\sum_{ijl\sigma s\lambda}\int dy d\veck 
   \int\frac{d\ub{p}}{\sqrt{16\pi^3p^+}}     \left\{h_{ijl}^{\sigma s\lambda}(y,\veck)
        a_{l\lambda}^\dagger(y,\veck;\ub{p})
           b_{js}^\dagger(1-y,-\veck;\ub{p})b_{i\sigma}(\ub{p}) \right. \nonumber \\
&& \left. \rule{2in}{0mm}
         +h_{ijl}^{\sigma s\lambda *}(y,\veck)b_{i\sigma}^\dagger(\ub{p})
     b_{js}(1-y,-\veck;\ub{p})a_{l\lambda}(y,\veck;\ub{p}) \right\},  \nonumber
\eea
with $\epsilon^\lambda=(-1,1,1,1)$ and the $h_{ijl}^{\sigma s\lambda}$
known vertex functions.

\section{The dressed-electron state}

The right and left-hand valence states ($\ob{\Pminus}$ is not Hermitian!) are
$|\phi_a^\pm\rangle=\sum_i z_{ai} b_{i\pm}^\dagger(\ub{P})|0\rangle$ and 
$\langle\tilde\phi_a^\pm|=\langle0|\sum_i \tilde{z}_{ai} b_{i\pm}(\ub{P})$.
We approximate the $T$ operator with the simplest form
\bea
T&=&\sum_{ijls\sigma\lambda}\int dy d\vec{k}_\perp 
   \int\frac{d\ub{p}}{\sqrt{16\pi^3}}\sqrt{p^+} t_{ijl}^{\sigma s\lambda}(y,\vec{k}_\perp)
 a_{l\lambda}^\dagger(yp^+,y\vec{p}_\perp+\vec{k}_\perp) \\
 && \rule{0.5in}{0mm} \times
   b_{js}^\dagger((1-y)p^+,(1-y)\vec{p}_\perp-\vec{k}_\perp)b_{i\sigma}(\ub{p}).
   \nonumber
\eea
The effective Hamiltonian $\ob{\Pminus}$ can then be constructed~\cite{LFCCqed}.
From this effective Hamiltonian, the right and left-hand
valence-sector equations become, for $a=0,1$,
\be
m_i^2 z_{ai}^\pm +\sum_j I_{ij} z_{aj}^\pm = M_a^2 z_{ai}^\pm\;\;\;\;
\mbox{and}\;\;\;\;
m_i^2 \tilde{z}_{ai}^\pm +\sum_j (-1)^{i+j}I_{ji} \tilde{z}_{aj}^\pm 
     = M_a^2 \tilde{z}_{ai}^\pm,
\ee
with $M_a$ the $a$th eigenmass and the self-energy given by
\be
I_{ji}=(-1)^i\sum_{i'ls\lambda}(-1)^{i'+l}\epsilon^\lambda
        \int \frac{dy d\veckp}{16\pi^3}
        h_{ji'l}^{\sigma s\lambda*}(y,\vec k_\perp) 
        t_{ii'l}^{\sigma s\lambda}(y,\vec k_\perp).
\ee
The valence eigenvectors are orthonormal and complete
in the following sense:
\be
\sum_i (-1)^i \tilde{z}_{ai}^\pm z_{bi}^\pm=(-1)^a \delta_{ab} \;\;
\mbox{and}\;\;
\sum_a (-1)^a z_{ia}^\pm \tilde{z}_{ja}^\pm = (-1)^i \delta_{ij}.
\ee

The $t$ functions satisfy the projection of the effective 
eigenvalue problem onto one-electron/one-photon states,
orthogonal to $|\phi\rangle$, which gives~\cite{LFCCqed}
\bea
\sum_i(-1)^i z_{ai}^\pm\left\{h_{ijl}^{\pm s\lambda}(y,\veck)
+\frac12 V_{ijl}^{\pm s\lambda}(y,\veck) 
+\left[\frac{m_j^2+k_\perp^2}{1-y}+\frac{\mu_{l\lambda}^2+k_\perp^2}{y}-m_i^2\right]
                 t_{ijl}^{\pm s\lambda}(y,\veck) \right. &&  \\
\left. +\frac12\sum_{i'} \frac{I_{ji'}}{1-y} t_{ii'l}^{\pm s\lambda}(y,\veck)
-\sum_{j'}(-1)^{i+j'}t_{j'jl}^{\pm s\lambda}(y,\veck)I_{j'i} \right\}=0, \nonumber
\eea
with the vertex correction
\bea
V_{ijl}^{\sigma s\lambda}(y,\veck)
&=&\sum_{i'j'l'\sigma' s\lambda'}(-1)^{i'+j'+l'}\epsilon^{\lambda'}
  \int \frac{dy' d\veckp}{16\pi^3}
  \frac{\theta(1-y-y')}{(1-y')^{1/2}(1-y)^{3/2}} \\
&&\times 
  h_{jj'l'}^{ss'\lambda'*}(\frac{y'}{1-y},\veckp+\frac{y'}{1-y}\veck)
  t_{i'j'l}^{\sigma's'\lambda}(\frac{y}{1-y'},\veck+\frac{y}{1-y'}\veckp)
  t_{ii'l'}^{\sigma\sigma'\lambda'}(y',\veckp).  \nonumber
\eea
To partially diagonalize in flavor, we define
%
%\be
$C_{abl}^{\pm s\lambda}(y,\veck)
  =\sum_{ij}(-1)^{i+j}z_{ai}^\pm \tilde{z}_{bj}^\pm t_{ijl}^{\pm s\lambda}(y,\veck)$.
%\ee
%
With analogous definitions for $H$, $I$, and $V$, we have
\bea
\lefteqn{\left[M_a^2-\frac{M_b^2+k_\perp^2}{1-y}-\frac{\mu_{l\lambda}^2+k_\perp^2}{y}\right]
   C_{abl}^{\pm s\lambda}(y,\veck)}&& \\
&&   =H_{abl}^{\pm s\lambda}(y,\veck)
   +\frac12\left[V_{abl}^{\pm s\lambda}(y,\veck)
      -\sum_{b'}\frac{I_{bb'}}{1-y}C_{ab'l}^{\pm s\lambda}(y,\veck)\right] \nonumber
\eea
to be solved simultaneously with the valence sector equations, which depend
on $C/t$ through the self-energy matrix $I$.
Notice that the physical mass $M_b$ has replaced the bare mass in
the kinetic energy term, without any need for
sector-dependent renormalization~\cite{SecDep}.

In order to compute matrix elements, such as appear in the computation
of form factors, we need the left-hand eigenstate.
The dual to $\langle\tilde\psi|=\sqrt{Z}\langle\psi|e^T$ is a right eigenstate of $\ob{\Pminus}^\dagger$
\be \label{eq:LHket}
|\widetilde\psi_a^\sigma(\ub{P})\rangle=|\widetilde\phi_a^\sigma(\ub{P})\rangle 
+\sum_{jls\lambda}\int dy d\veck\sqrt{\frac{P^+}{16\pi^3}}
l_{ajl}^{\sigma s\lambda}(y,\veck)a_{l\lambda}^\dagger(y,\veck;\ub{P}) 
b_{js}^\dagger(1-y,-\veck;\ub{P})|0\rangle,
\ee
The flavor-diagonal left-hand wave functions are
%
%\be
$D_{abl}^{\pm s\lambda}(y,\veck)\equiv\sum_j(-1)^j z_{bj}^s l_{ajl}^{\pm s\lambda}(y,\veck)$.
%\ee
%
They satisfy the coupled equations~\cite{LFCCqed}
\bea
\lefteqn{\left[M_a^2-\frac{M_b^2+k_\perp^2}{1-y}
    -\frac{\mu_{l\lambda}^2+k_\perp^2}{y}\right]
        D_{abl}^{\sigma s\lambda}(y,\veck)}&& \\
 && =\tilde{H}_{abl}^{\sigma s\lambda}(y,\veck)
 +W_{abl}^{\sigma s\lambda}(y,\veck)
     -\sum_{b'} J_{b'a}^\sigma \tilde{H}_{b'bl}^{\sigma s\lambda*}(y,\veck),
     \nonumber
\eea
where $W_{abl}^{\sigma s\lambda}$ is a vertex-correction analog of
$V_{abl}^{\sigma s\lambda}$, though linear in $D$, and $J_{ba}^\sigma$
is a self-energy analog of $I_{ba}$.
Solutions for $M_a$, $z_{ai}^\sigma$, $\tilde{z}_{ai}^\sigma$, 
and $C_{abl}^{\sigma s \lambda}$ are used as input.

\section{Anomalous magnetic moment}

We compute the anomalous moment $a_e$ from the spin-flip
matrix element~\cite{BrodskyDrell} of the current 
$J^+=\overline{\psi}\gamma^+\psi$ coupled to a photon of momentum $q$
in the Drell--Yan ($q^+=0$) frame~\cite{DrellYan}
\be
16\pi^3\langle\psi_a^\sigma(\ub{P}+\ub{q})|J^+(0)|\psi_a^\pm(\ub{P})\rangle
=2\delta_{\sigma\pm}F_1(q^2)\pm\frac{q^1\pm iq^2}{M_a}\delta_{\sigma\mp}F_2(q^2).
\ee
In the limit of infinite Pauli--Villars masses,
and with $M_0=m_e$, the electron mass, we find~\cite{LFCCqed}
\bea
F_1(q^2)&=&\frac{1}{\cal N}\left[1+\sum_s\int\frac{dy d\veck}{16\pi^3}
\left\{\sum_{\lambda=\pm} 
  l_{000}^{\pm s\lambda *}(y,\veck-y\vec{q}_\perp)t_{000}^{\pm s\lambda}(y,\veck)
                      \right.\right.\\
 && \left. \left. \rule{1.5in}{0mm}
 -\sum_{\lambda=0}^3\epsilon^\lambda
    l_{000}^{\pm s \lambda *}(y,\veck) t_{000}^{\pm s \lambda}(y,\veck)\right\}\right]
 \nonumber
\eea
and
\be
F_2(q^2)=\pm\frac{2m_e}{q^1\pm iq^2}\frac{1}{\cal N}
\sum_s\sum_{\lambda=\pm}\int\frac{dy d\veck}{16\pi^3} 
    l_{000}^{\mp s\lambda *}(y,\veck-y\vec{q}_\perp)
                         t_{000}^{\pm s\lambda}(y,\veck),
\ee
with
\be
{\cal N}=1-\sum_s\sum_{\lambda=0,3}\epsilon^\lambda
   \int\frac{dy d\veck}{16\pi^3} l_{000}^{\pm s\lambda*}(y,\veck)
          t_{000}^{\pm s\lambda}(y,\veck).
\ee
A second term is absent in $F_2$ because $l$ and $t$ are orthogonal
for opposite spins.  The $q^2\rightarrow0$ limit can be taken, 
to find $F_1(0)=1$ and
\be
a_e=F_2(0)=\pm m_e\sum_{s\lambda}\epsilon^\lambda
\int \frac{dy d\veck}{16\pi^3}
y l_{000}^{\mp s\lambda *}(y,\veck)
\left(\frac{\partial}{\partial k^1}\mp i\frac{\partial}{\partial k^2}\right)
t_{000}^{\pm s\lambda}(y,\veck).
\ee
As a check, we can consider a perturbative solution 
\be
t_{000}^{\sigma s\lambda}
=l_{000}^{\sigma s\lambda}
=h_{000}^{\sigma s\lambda}
/\left[m_e^2-\frac{m_e^2+k_\perp^2}{1-y}
    -\frac{\mu_{l\lambda}^2+k_\perp^2}{y}\right].
\ee
Substitution into the expression for $a_e$ gives
immediately the Schwinger result~\cite{Schwinger} $\alpha/2\pi$,
in the limit of zero photon mass, for any covariant gauge.

\section{Summary}

The LFCC method provides a nonperturbative approach to
bound-state problems in quantum field theories without
truncation of the Fock space and without the uncanceled 
divergences and spectator dependence that such truncation
can cause.  The approximation is instead a truncation
of the operator $T$ that generates contributions from 
higher Fock states.  It is systematically improvable
through the addition of more terms to $T$, with increasing
numbers of particles created and annihilated.

To complete the application to the dressed-electron state,
we need to solve numerically the coupled systems that
determine the $t$ and $l$ functions and to use these
solutions to compute the anomalous moment.  Within
the arbitrary-gauge formulation, we can test directly
for gauge dependence~\cite{ArbGauge}.  A more complete
investigation of QED would include consideration of
the dressed-photon state, contributions from electron-positron
pairs to the dressed-electron state, and true bound states
such as muonium and positronium.  These will provide
some guidance for applications to quantum chromodynamics,
particularly in extensions of the holographic model for
mesons~\cite{hQCD}.

\acknowledgments
This work was done in collaboration with 
J.R. Hiller and supported in part by
the US Department of Energy and
the Minnesota Supercomputing Institute.

\end{document}